\documentclass[aps]{revtex4}%
\usepackage{amsfonts}
\usepackage{amsmath}
\usepackage{amssymb}
\usepackage{graphicx}%
\providecommand{\U}[1]{\protect\rule{.1in}{.1in}}
\begin{document}
\preprint{ }
\title{Optimization in the Parikh-Wilczek tunneling model of Hawking radiation for Kerr-Newman Black Holes}

\author{Auttakit Chatrabhuti$^{1, 2}%
$\footnote{Email: auttakit@sc.chula.ac.th} and Khem Upathambhakul$^{1}$\footnote{Email: keimatham@gmail.com}}
\affiliation{$^{1}$Particle Physics Research Laboratory, Department of Physics, Faculty of Science, Chulalongkorn University, Bangkok 10330,
Thailand}
 \affiliation{$^{2}$Thailand Center of Excellence in Physics, CHE, Ministry of Education, Bangkok 10400, Thailand}

\begin{abstract}
In this short report, we investigate the mutual information hidden in the Parikh-Wilczek tunneling model of Hawking radiation for Kerr-Newman black holes. By assuming the radiation as an optimization process, we discuss its effect on time evolution of rotating (charged and uncharged) black holes.  For uncharged rotating black holes evaporating under the maximum mutual information optimization, their scale invariant rotation parameter $a_*=a/M$ is almost constant at the early stage but rapidly increase at the very last stage of the evaluation process.  The value of rotation parameter at the final state of evaporation depends on the initial condition of the black hole.  We also found that the presence of electric charge can cause the black holes lose their angular momentum more rapidly than they lose mass.  The charged-rotating black holes asymptotically approach a state which is described by $a_*= 0$ and $Q/M = 1$.
\end{abstract}
\date{\today}
\maketitle

\section{Introduction}

While the conventional framework of Hawking radiation based on perturbation in a fixed gravitational background with horizons plays an important role in black hole physics, its pure thermal emission spectrum leads to the information loss paradox.   In 2000, Parikh and Wilczek \cite{Parikh} proposed a semi-classical framework to implement Hawking radiation as a tunneling process.  They demonstrated that the emission spectrum of black hole radiation is not strictly pure thermal when the energy conservation is enforced.  Their result is consistent with an underlying unitary theory and information is conserved.  For the Schwarzschild black hole of mass $M$, the tunneling probability for emission of energy $\omega$ is $\Gamma(M,\omega) \sim \exp\{-8\pi\omega(M-\omega/2)\}$ in the natural units and is related to the change of the black hole's entropy.    One can show that this radiation is not thermal by considering sequential emissions of energies $\omega_1$ and $\omega_2$ i.e. by showing that two consecutive emissions are not independent $\Gamma(M,\omega_1+\omega_2) \neq \Gamma(M,\omega_1)\cdot\Gamma(M,\omega_2)$.  The tunneling probability for the second emission should be denoted as the condition possibility $\Gamma(M, \omega_2|\omega_1)\equiv \Gamma(M-\omega_1,\omega_2) = \exp\{-8\pi\omega_2(M-\omega_1-\omega_2/2)\}$ and it clearly depends on $\omega_1$ such that $\Gamma(M,\omega_2|\omega_1)\neq\Gamma(M,\omega_2)$.    The amount of correlation inside Hawking radiation can be measured by the mutual information between the two consecutive quantum emissions defined by \cite{Cai}
\begin{equation}
S_{MI}(M,\omega_2:\omega_1) \equiv S(M,\omega_2|\omega_1)-S(M,\omega_2),
\end{equation}
where the entropy function $S(M,\omega_2) = -\ln \Gamma(M,\omega_2)$ and the condition entropy $S(M,\omega_2|\omega_1) = -\ln \Gamma(M,\omega_2|\omega_1)$.   One can see that that the mutual information vanishes by definition if two emissions are independent.    For the tunneling model of Schwarzschild black hole, we obtain $S_{MI}(M,\omega_2:\omega_1) = 8\pi\omega_1\omega_2$.  As a black hole evaporates, it entropy monotonically decreases.   By taking into account the fact that total entropy must be conserved in unitary theory, it is natural to assume that the mutual information stored in each pair of consecutive emissions is nonnegative.  For the Schwarzchild black hole, nonnegativity of the mutual information simply requires the positivity of mass or energy of each emission.   Recently, Kim and Wen investigated the mutual information in the case of Reissner-Norstr\"om black hole \cite{Wen} and found that the nonnegativity condition of mutual information enforces bounds for charge-mass ratio of emitted particles.  They showed that the upper bound is between $\sqrt{2}$ and $1$.  The authors in \cite{Wen} also proposed that Hawking radiation can be understood as an optimization process.  Since, the emission carrying more mutual information has more probability.  The emission with maximum mutual information (MMI) dominates the process.  From the information point of view, the charge black hole radiates most efficiently by giving away as much as information as possible by radiating with the optimized charge-mass ratio.  The evolution of charged black hole evaporating with respect to the MMI optimization was also studied.  It was claim that the nonnegative mutual information can serve as an underlying principle governing dynamics of the black hole evaporation under MMI optimization.  In this paper, we would like to explore this idea further in the case of rotating charged Kerr-Newman black holes.  The fate of Kerr black hole was discussed in great detail by \cite{Chambers,Taylor,Page} and it was claimed that  a black hole emitting solely scalar radiation will approach a final asymptotic state with non-zero angular momentum.  It is interesting to see how the optimization process dictates the evolution of rotating black holes. 

Let us start by considering the Kerr-Newman Black hole with mass $M$, Charge $Q$ and angular momentum $J = M a$, we can write the entropy function as \cite{Zhang, Zhu}
\begin{eqnarray}
S(M,Q,J; w, q, m) &=& -2\pi\left\{M^2 - (M-\omega)^2 + M\sqrt{M^2-a^2-Q^2}-\frac{1}{2}\left[Q^2-(Q-q)^2 \right]\right. \nonumber\\
&& \left.-(M-\omega)\sqrt{(M-\omega)^2-\frac{(Ma-j)^2}{(M-\omega)^2}  - (Q-q)^2}\right\},
\label{entropy function}
\end{eqnarray}
for each emission of a particle with energy (or mass) $\omega$, charge $q$ and angular momentum $j$.   The factor $\frac{(Ma-j)^2}{(M-\omega)^2} $ in  (\ref{entropy function}) is the result of angular momentum conservation.  Note that in the case of rotating black hole, tunneling particles might contain angular momentum.  Thus, we must consider not only the energy and charge conservation but also the angular momentum conservation.   Total angular momentum of the system is given by black hole angular momentum before emission, $J=Ma$.  After emitting a particle of mass $\omega$, charge $q$ and angular momentum $j$, black hole state has changed from $(M,Q, J)$ to $(M-\omega,Q-q,J-j)$.  In this new state, the specific angular momentum $a$ should be replaced by 
\begin{equation}
a = \frac{J}{M} \rightarrow a^\prime = \frac{J-j}{M-\omega} = \frac{Ma-j}{M-\omega}.
\label{a_def}
\end{equation}
Note that the authors in \cite{Zhang} assumed that the specific angular momentum $a$ is constant (or equivalently $j = \omega a$) during the evaporation process.  This choice is quite unnatural because it will make the scale invariant rotation parameter $a_* = a/M$ monotonically increases and exceeds the extremal limit at a finite time steps.  So, we leave $j$  to have any arbitrary value at the moment (as in \cite{Zhu}).  However, as we will see later on, there exist the optimized value for $j/\omega^2$ which maximize the information giving off by Hawking radiation.   

\begin{figure}[ptb]
\begin{center}
\includegraphics[scale=.7 ]{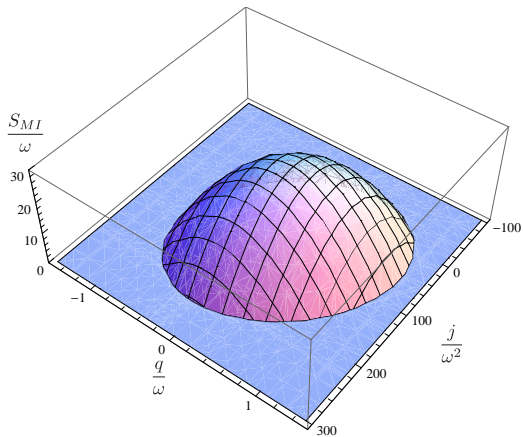}
\end{center}
\caption{The mutual information of Kerr-Newman black hole with $\omega/M=10^{-2}$, $a_* = 0.4$ and $Q/M = 0.6$ is shown as a function of the charge-mass ratio $q/\omega$ and angular momentum-mass square ratio $j/\omega^2$ of the emission.  We also plot the region of negative ratio in order to locate the maximum value of $S_{MI}$.}%
\label{fig1}%
\end{figure}

The mutual information between two consecutive emission of mass $\omega_1$, $\omega_2$, charge $q_1$, $q_2$ and angular momentum $j_1$, $j_2$ can be defined as
\begin{equation}
S_{MI}(M,Q,J;\omega_2,q_2,j_2:\omega_1,q_1,j_1) \equiv S(M,Q,J;\omega_2,q_2,j_2|\omega_1,q_1,j_1) - S(M,Q,J;\omega_2,q_2,j_2),
\label{SMI}
\end{equation}
where the condition entropy for Kerr-Newman case is $S(M,Q,J;\omega_2,q_2,j_2|\omega_1,q_1,j_1) \equiv S(M-\omega_1,Q-q_1,J-j_1;\omega_2,q_2,j_2)$.   This quantity has nontrivial dependent on $M$, $Q$ and $a$.  For the limit of large black hole mass, large charge and/or large angular momentum, we take  $M \gg \omega_1, \omega_2$, $|Q| \gg |q_1|, |q_2|$, $J\gg j_1, j_2$ and $M>|Q|,a$.   The mutual information takes the simple form 
\begin{equation}
    S_{MI}(M,Q,J;\omega_2,q_2,j_2:\omega_1,q_1,j_1) = 8\pi\omega_1\omega_2 - 4\pi q_1q_2 + O(\omega/M).
\end{equation}
This is the same as the large black hole mass and charge limit for Reissner-Norstr\"om black hole \cite{Wen}.  For simplicity, we now assume that the two emission are identical, i.e.   $\omega_1=\omega_2 = \omega$, $q_1=q_2=q$ and $j_1=j_2=j$.   The above large mass limit gives us an upper bound for the charge and mass ratio $|q|/\omega \leq \sqrt{2}$.   In Figure \ref{fig1}, we plot the mutual information between two consecutive emission as a function of the charge-mass ratio and the angular momentum-mass ratio of the emission.  As seen in the figure, the nonnegativity of $S_{MI}$ enforce a bound for allowed value of $q/\omega$ and $j/\omega^2$.  

Note that monotonically decreasing of black hole entropy is essential for the nonnegativity of $S_{MI}$.  However in the conventional framework of  Hawking radiation, the black hole entropy is not alway monotonically decreasing, especially in the case of near extremal Kerr black holes ($a_* \sim 1$) as discussed in \cite{Taylor}.  In \cite{Page}, Page explained that this behavior can be understood as the effect of superradiance modes:  At high values of $a_*$, the emission is primarily the spontaneous emission corresponding to the stimulated emission of superradiant scattering. In this process, pairs are created in the ergosphere with particles being emitted to infinity with positive energies and their antiparticles going down the black hole with negative energies as measured at infinity but positive energies as measured locally.  Thus heat flows down the hole as well as out to infinity, increasing the entropy of both black hole and the outside during the early stage of Hawking radiation.   This behavior obviously involves irreversible and non-unitary processes due to heat flow and increasing of total entropy.  On the other hand, in the Parikh-Wilczek tunneling framework, the total entropy is constant so that the information is conserved.  Such non-unitary process is not allowed \cite{Hu}.  Thus we will assume that the black hole entropy alway decrease and the nonnegativity condition for $S_{MI}$ is still hold in the tunneling model for Hawking radiation.

\section{Optimization in tunneling model for Kerr-Newman black hole}

The optimized $|q|/\omega$ and $j/\omega^2$ ratios of emission with MMI are denoted by $\gamma_M^q$ and $\gamma_M^j$ respectively.  If black hole mass $M$, charge $Q$ and specific angular momentum $a$ are given, they can be obtained by solving non-linear equations $\partial(S_{MI})/\partial(q/\omega) = 0$ and $\partial(S_{MI})/\partial(j/\omega^2)=0$.  Although it is difficult to find exact solutions for these equation, their asymptotic expansions can give us some insight on qualitative behavior of the evaporation process.  By assuming two consecutive emissions are small, we obtain the following asymptotic solutions: 
\begin{itemize}
  \item  \textbf{The First Asymptotic solution (AS-I):}\\
  The first solution gives us charge-mass ratio and angular momentum-mass ratio of MMI as 
  \begin{equation}
       \gamma_{M}^q  =  0\;\text{   and   }\; \gamma_{M}^j = \frac{a}{\omega} \left( \frac{2M^2-Q^2}{M^2-Q^2}\right). 
  \label{ASI}
  \end{equation}
  In this limit, the charge-mass ratio of emission vanishes.  Black hole emissions carry off energy and angular momentum but no charge.    
  
  \item \textbf{The Second Asymptotic solution (AS-II):}\\
  The second solution gives us charge-mass ratio and angular momentum-mass ratio of MMI as
     \begin{equation}
       \gamma_{M}^q  =  \frac{Q(Q^2+2a^2)}{M(M^2-a^2)+(M^2-Q^2-a^2)^{3/2}}\;\text{   and  } \; \gamma_{M}^j =  0. 
  \label{ASII}
  \end{equation}  
  In this limit, the angular momentum-mass ratio of emission vanishes.  The Hawking radiation carry off energy and charge but no angular momentum.  For the Reissner-Norstr\"om limit where $a=0$, this solution reduce to the charge-mass ratio presented in \cite{Wen}.   
  \end{itemize}
  
\begin{figure}[ptb]
\begin{center}
\includegraphics[scale=.7 ]{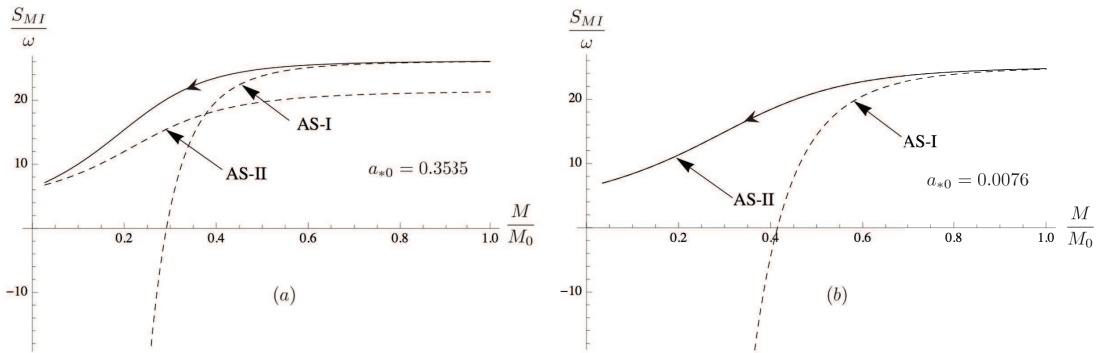}
\end{center}
\caption{The mutual information between two consecutive emissions is plotted versus the fractional mass of a Kerr-Newman black hole evaporating under under MMI optimization.  The two dash-lines represent evolution paths with respect to asymptotic solution AS-I in (\ref{ASI}) and AS-II in (\ref{ASII}).  Numerical solution is shown by the solid line.  In figure (a), a black hole starts with an initial state: $a_{*0}= 0.3535$ ($\omega/M_0 = 10^{-5}$, $\rho_0 = 0.5$, $\theta_0 = \pi/4$).  At the beginning, the evaporation process starts very close to AS-I.  As the mutual information $S_{MI}/\omega$ of AS-I decreasing, the evolution line moves away from the first asymptotic solution and approaches the second asymptotic solution AS-II which has higher mutual information at the later stages of the evaporation.  While in figure (b)  a black hole starts with very small initial rotation parameter, $a_{*0}= 0.0078$ ($\omega/M_0 = 10^{-5}$, $\rho_0 = 0.5$, $\theta_0 = 0.001\pi/2$).   We can see that AS-II has higher mutual information $S_{MI}/\omega$ and dominates the whole evaporation process.  }%
\label{fig2}%
\end{figure}

For simplicity, we classify black hole initial states of mass $M_0$, charge $Q_0$ and specific angular momentum $a_0$ by defining scale invariant parameters 
\begin{equation}
\rho = \frac{\sqrt{Q_0^2 + a_0^2}}{M_0}\; \text{     and    }\; \theta_0 = \arctan \left( \frac{a_0}{|Q_0|}\right).
\end{equation}
Note that the extremal limit can be achieved by setting $\rho =1$.  By setting $\theta_0 = 0$, we obtain the Reissner-Norstr\"om black hole.  While, $\theta_0 = \pi/2$ limit gives us the Kerr black hole.   It is also useful to classify black holes states by their scale invariant rotation parameter $a_* = \frac{J}{M^2}=\frac{a}{M}$.

At large value of $a_*$,  AS-I dominates the early stage of evaporation process.  However, its mutual information decreases very fast and becomes negative.   Thus  AS-II becomes the dominant contribution at the later stage.  On the other hand for a Kerr-Newman black hole with small $a_*$, the second asymptotic solution dominates the whole process.   These behaviors are numerically demonstrated in Figure \ref{fig2} where $S_{MI}$ is plotted versus the fractional mass $M/M_0$.  The two dash-lines represent evolution paths with respect to asymptotic solution AS-I in (\ref{ASI}) and AS-II in (\ref{ASII}).  Numerical solution is shown by the solid-line.   In figure (a) we consider a non-extremal Kerr-Newman black hole with initial state $a_{*0}= 0.3535$ ($\omega/M_0 = 10^{-5}$, $\rho_0 = 0.5$, $\theta_0 = \pi/4$).   We can see that at the early stage the mutual information due to emissions under AS-I is greater  than those under AS-II path.  Thus, the hole starts its evaporation process very close to the AS-I curve.  However, at the later stage, the mutual information for AS-I decreases much faster and becomes lower than those of AS-II.  Since the hole tries to release as much information as possible, it evolves farther away from AS-I and approaches the AS-II curve at the later stage.  On the other hand, figure (b) shows the evolution of $S_{MI}$ for a black hole evaporating under MMI with the initial state $a_* = 0.0078$ ($\omega/M_0 = 10^{-5}$, $\rho_0 = 0.5$, $\theta_0 = 0.001\pi/2$).  In this case, evaporation near AS-II dominates the whole process and emissions with zero-angular momentum become dominant.  Note that this seems roughly to agree with previous studies in the case of Kerr black holes by Page \cite{Page} and by Chambers, Hiscock and Taylor \cite{Chambers} where they found that, at low $a_*$, dominant modes for scalar emission would carry off no angular momentum.  In their model, as $a_*$ increases, the superradiant effect becomes more effective and increases the angular momentum loss rate \cite{Chambers,Taylor}.

\begin{figure}[ptb]
\begin{center}
\includegraphics[scale=.7 ]{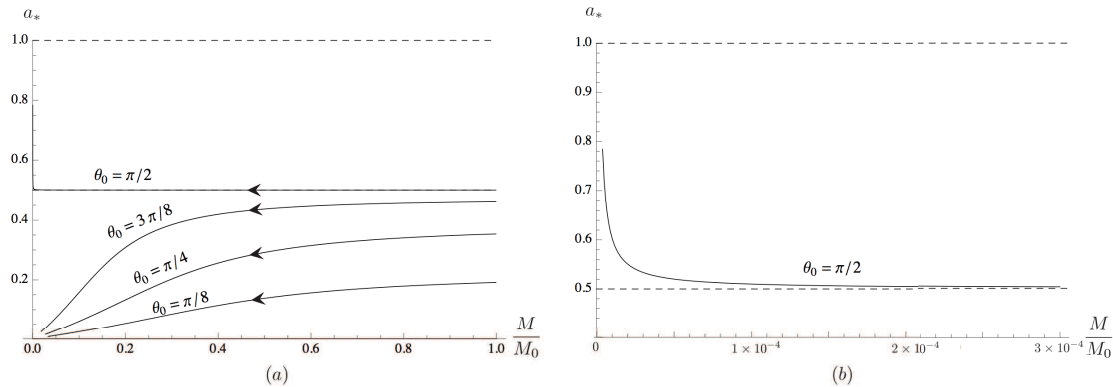}
\end{center}
\caption{(a) The fractional mass of a black hole evaporating under MMI optimization is plotted versus its scale invariant rotation parameter, $a_*= J/M^2$, for initial states with $\omega/M_0 = 10^{-5}$, $\rho_0 = 0.5$, $\theta_0 = \pi/8, \pi/4, 3\pi/8$, and $\pi/2$.  For Kerr black hole represented by $\theta_0=\pi/2$ line, $a_*$ stays at roughly its initial value $\rho_0$ except for the very last stage of the evaporation process.  In contrast rotating black holes with non-vanishing charge evolve to final states with very small value of $a_*$. (b) The behavior of a Kerr black holes at the final stage of the evaporation process is shown.  As seen in the figure, $a_*$ rapidly increases and reach the final value $a_{*f} \sim 0.7842$ at the end of the evaporation.}%
\label{fig3}%
\end{figure}

In Figure \ref{fig3}, the fractional mass of a black hole is plotted versus its scale invariant rotation parameter, $a_*$, for initial states with $\omega/M_0 = 10^{-5}$, $\rho_0 = 0.5$, $\theta_0 = \pi/8, \pi/4, 3\pi/8$, and $\pi/2$.    There are two important features that we would like to draw your attention to.   Firstly, in the case of a Kerr black hole ($\theta_0 = \pi/2$), $a_*$ seems to be unchanged and equal to its initial value $a_{*0} = 0.5000$ for most of the evaporation process.  However, at the final stage, its value increases very rapidly and approaches a finite value $a_{*f} \sim 0.7842$.  Secondly, in the case of rotating black holes with non-zero charge, their parameter $a_*$ decrease and are close to zero value as black holes approaching to their final states.   

The first behavior  can be understood analytically.  Recall that after each emission the rotation parameter of a Kerr black hole is changed from $a_*= a/M$ to $a_*^\prime = a^\prime/(M-\omega)$.   In the case of Kerr black hole, we can solve the equation $\partial(S_{MI})/\partial(j/\omega^2) = 0$ and obtain the exact solution
\begin{equation}
\gamma_M^{j,\text{Kerr}} = \frac{a}{\omega}\left(\frac{2-\omega/M}{1-2\omega^2/M^2} \right).
\label{Kerr}
\end{equation}
From (\ref{Kerr}), the angular momentum carried off by the emission is $j=\gamma_M^{j,\text{Kerr}} \omega^2$.  By using definition of $a^\prime$ in (\ref{a_def}) and assuming the emission is small, we can show that $\Delta a_* \equiv  a_*^\prime - a_* \sim O(\omega/M)^2$.  This implies that the parameter $a_*$ is almost constant when $M \gg \omega$.  As the black hole approaches its final state, its mass is roughly as the same order as emission energy  $M \gtrsim \omega$ and we obtain
\begin{equation}
a^\prime_* = a_* \left(\frac{M^2}{M^2-2\omega^2} \right).
\label{final state}
\end{equation}
We can show that $\Delta a_*/a_* = 2\omega^2/(M^2-2\omega^2) >0$.  This implies $a_*$ increases at the final stage of the evaporation process.  
%For large $a_{*}$, we can estimate the value of the scale invariant rotation parameter of the final state $a_{*f}$ by setting $a^\prime_* =1$ and $a_* = a_{*f}$ in (\ref{final state}).  $a_{*f} \approx \frac{M_f^2-2\omega^2}{M_f^2}$.  For our example considered in Figure \ref{fig3}, the mass of black hole near its final state is $M_f \approx 3 \omega$.   This approximation give us $a_{*f} \approx 0.7778$ which is close to the numerical value from numerical simulation $a_{*f} = 0.7842$.  
Note that this result is different from the fate of Kerr black holes previously proposed in \cite{Chambers, Taylor} where it was claimed that black holes emitting solely scalar radiation will approach a final asymptotic states with  $a_{*f} \sim 0.555$.   In our framework, the rotation parameter of the final state depends on the initial value $a_{*0}$.  For example, the Kerr black hole with $a_{*0} = 0.9000$ approaches its final state with $a_{*f} \sim 0. 9923$.

In order to understand the second behavior, it is essential to know how the black hole lose their charge during the evaporation process.  In Figure 4, we consider black holes with initial states $\omega/M_0 = 10^{-5}$, $\rho_0 = 0.5$, $\theta_0 = 0, \pi/8, \pi/4$ and $3\pi/8$.  The black hole scale invariant charge parameter $Q/M$ is plotted versus its fractional mass.  It appears that the black hole's charge loss rate is not high enough to exceed its mass loss rate.  This causes the charge parameter to increase.  We can see that non-vanishing charge black holes evolve to final states with $Q/M$ close to unity.  Recall that the extremal constraint can be written in terms of the charge and rotation parameter as $Q^2/M^2+a_*^2 \leq 1$.  As $Q/M$ increases and becomes close to unity at the final state, it is natural to expect $a_*$ to decrease and approach zero at the later stage.   Moreover, since AS-I dominates for most of the evaporation process,  emissions carry off only angular momentum are dominant.  The angular momentum loss rate is much higher than the charge loss rate until the final stage of evaporation.

In Figure \ref{fig5}, the square of mass charge parameter $Q^2/M^2$ is shown plotted versus the rotation parameter square.  We can see that non-extremal Kerr-Newman black holes evolve to the final states close to extremal Reissner-Norstr\"om black holes ($Q/M \approx 1$, $a_* \approx 0$).  As discuss in \cite{Wen}, the mutual information decreases with $a_*$.  The closer the black hole approach the extremal limit the less information is carried off by radiation.  The evaporation process is expected to stop at the extremal limit where mutual information vanishes.  However, non-extremal black holes might need infinite time steps until they become extremal \cite{Fabbri}. 

\begin{figure}[ptb]
\begin{center}
\includegraphics[scale=.7 ]{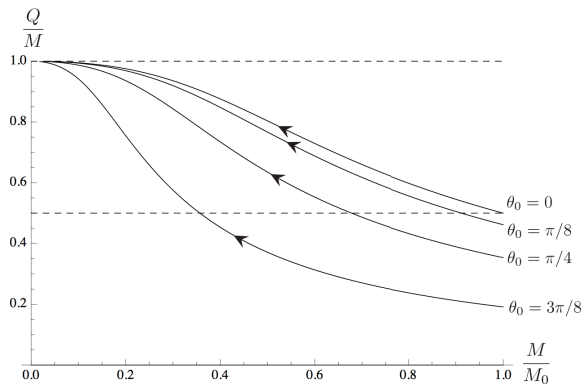}
\end{center}
\caption{The fractional mass of a black hole evaporating under MMI optimization is shown plotted versus its scale invariant charge parameter, $Q/M$, for initial states with $\omega/M_0 = 10^{-5}$, $\rho_0 = 0.5$, $\theta_0 = 0, \pi/8, \pi/4$, and $3\pi/8$.  The initial state with $\theta_0 = 0$ represents the Reissner-Norstr\"om black hole.  All initial states evolve to final states with $Q/M \approx 1$.}%
\label{fig4}%
\end{figure}

\begin{figure}[ptb]
\begin{center}
\includegraphics[scale=.7 ]{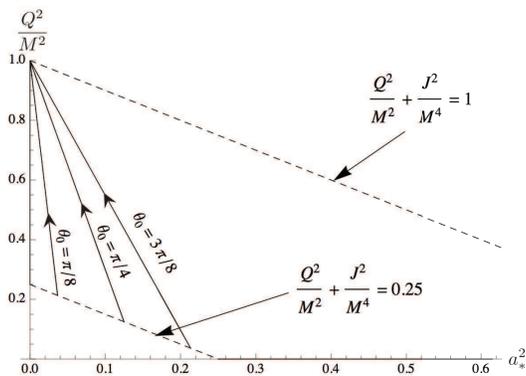}
\end{center}
\caption{Evaporation under the MMI optimization with initial states $\omega/M_0 = 10^{-5}$, $\rho_0 = 0.5$, $\theta_0 = \pi/8, \pi/4$, and $3\pi/8$ are considered.  The square of the scale invariant charge parameter of a Kerr-Newman black hole $Q^2/M^2$ is plotted versus its scale invariant rotation parameter square $a_*^2$.  The upper dash-line represent the extremal limit.  The process lead to an final states very close to extremal limit with $Q/M \approx 1$ and $a_* \approx 0$.}%
\label{fig5}%
\end{figure}

\section{Discussion}

With the Parikh-Wilczek tunneling framework, we investigate the evolution of Kerr-Newman black holes evaporating under MMI optimization.  For the Kerr black holes ($Q=0$), their scale invariant rotation parameter $a_*=a/M$ is almost constant until the very last stage of the evaporation process.   As the Kerr black hole approaches its final state, its $a_*$ rapidly increases and approaches the finite value $a_{*f}$.  The value of $a_{*f}$ depends on the initial conditions of the black hole.  This result is in contrast with previous studies by the conventional framework \cite{Chambers, Taylor} where it was claim that the Kerr black hole approach asymptotic final state with $a_*\sim 0.555$.   We also found that the presence of electric charge can cause the black holes lose their angular momentum more rapidly than they lose mass.  The charged rotating black holes asymptotically approach a final state which is very close to Reissner-Norstr\"om black hole limit with $a_*=0$ and $Q/M = 1$.  

Finally, there are some interesting issues that we would like to mention here.  Firstly, the Parikh-Wilczek framework treats the emission process as reversible process and assumes the black hole and the outside approach an thermal equilibrium.  However, this thermal equilibrium can be unstable \cite{Hu}.  Moreover, we assume the entropy of black hole monotonically decrease but, as we mentioned earlier, this is in contrast with the conventional framework where both the black hole entropy and the total entropy can increase for high value of $a_*$ due to superradiance effect.  While such non-unitary process is not allowed in our model, it is interesting to consider superradiation phenomena in the Parikh-Wilczek tunneling framework.   Secondly, investigating black hole evaluation under the MMI optimization framework is more convenient than its conventional counterpart.  One just needs to know the correct form of the entropy function.  This new framework may provide us some qualitative understanding on the evaporation of higher dimensional black objects such as the five-dimensional black ring where investigation in the conventional framework has proved difficult.  We leave these issues for future investigations.

\section*{Acknowledgment}

The authors would like to acknowledge helpful discussions with Wen-Yu Wen.   We also thank Ahpisit Ungkitchanukit for valuable comments on the manuscript.  A.C. is supported by Thailand Toray Science Foundation.

%%%%%%%%%%%%%%%

\end{document}